\documentclass[10pt,twocolumn,letterpaper]{article}

\usepackage{iccv}
\usepackage[accsupp]{axessibility}  
\usepackage{booktabs}
\usepackage{times}
\usepackage{epsfig}
\usepackage{graphicx}
\usepackage{amsmath}
\usepackage{amssymb}
\usepackage{subfig}
\usepackage{listings}
\usepackage{makecell}
\usepackage{multirow}
\usepackage[dvipsnames]{xcolor}
\usepackage{pifont}
\newcommand{\cmark}{\ding{51}}%
\newcommand{\xmark}{\ding{55}}%

\usepackage{listings}
\usepackage{xcolor}

\definecolor{codegreen}{rgb}{0,0.6,0}
\definecolor{codegray}{rgb}{0.5,0.5,0.5}
\definecolor{codepurple}{rgb}{0.58,0,0.82}
\definecolor{backcolour}{rgb}{0.95,0.95,0.92}

\lstdefinestyle{mystyle}{
    backgroundcolor=\color{backcolour},   
    commentstyle=\color{codegreen},
    keywordstyle=\color{magenta},
    numberstyle=\tiny\color{codegray},
    stringstyle=\color{codepurple},
    basicstyle=\ttfamily\footnotesize,
    breakatwhitespace=false,         
    breaklines=true,                 
    captionpos=b,                    
    keepspaces=true,                 
    numbers=left,                    
    numbersep=5pt,                  
    showspaces=false,                
    showstringspaces=false,
    showtabs=false,                  
    tabsize=2
}

\usepackage[breaklinks=true,bookmarks=false]{hyperref}

\iccvfinalcopy 

\def\iccvPaperID{3} 

\ificcvfinal\fi
\begin{document}

\title{FSER: Deep Convolutional Neural Networks for Speech Emotion Recognition}

\author{Bonaventure F. P. Dossou\\
Jacobs University, Germany\\
{\tt\small f.dossou@jacobs-university.de}
\and
Yeno K. S. Gbenou\\
Drexel University, USA\\
{\tt\small  yg428@drexel.edu}
}

\maketitle

\begin{abstract}
Using mel-spectrograms over conventional MFCCs features, we assess the abilities of convolutional neural networks to accurately recognize and classify emotions from speech data. We introduce FSER, a speech emotion recognition model trained on four valid speech databases, achieving a high-classification accuracy of 95,05\%, over 8 different emotion classes: anger, anxiety, calm, disgust, happiness, neutral, sadness, surprise. On each benchmark dataset, FSER outperforms the best models introduced so far, achieving a state-of-the-art performance. We show that FSER stays reliable, independently of the language, sex identity, and any other external factor. Additionally, we describe how FSER could potentially be used to improve mental and emotional health care and how our analysis and findings serve as guidelines and benchmarks for further works in the same direction.
\end{abstract}

\section{Introduction}
Emotions are integral parts of daily communication between humans. Whether oral or written, properly understanding the interlocutor is key for a good communication. Even though emotions can be easily concealed in written communication, several psychological studies have shown that it is more difficult for humans to hide their feelings physically or vocally. More generally, facial expressions and voice tones are very good indicators of one's emotional state. Many psychological and physiological studies \cite{psy_3, psy_2, psy_4, psy_6, psy_1, psy_5} (not limited to the ones cited here) have proved, that emotions make us feel and act; stimulating and influencing both our facial expressions and voice tone. For instance, adrenaline is released in fearful situations to help us run away from danger, as excitement or joy can be expressed while we are talking with friends, family, cuddling our pets or practicing risky sports or activities such as mountaineering or skydiving.

The recognition of emotional states in speech, so called Speech Emotion Recognition (SER) is not a new concept in the fields of Artificial Intelligence and Machine Learning, but still is a very challenging task, that deserves more attention.

Convolutional neural networks (CNNs) are types of artificial neural networks, one of the most popular that have helped make breakthroughs in face recognition, segmentation and object recognition, handwriting recognition, and of course speech recognition. CNNs get their name from a well-known mathematical operation called convolution, and in most cases are used for efficient patterns recognition in data, notably including accurate images classification.\raggedbottom
\paragraph{Structure of the paper:} In sections \ref{dataset} and \ref{data_processing}, we respectively describe the speech corpora used in this study, and the preprocessing pipeline. In section \ref{results}, after introducing our model architecture, we describe our experiments, results and provide a use case of FSER. The last section is dedicated to the conclusion.
\section{Datasets, and Emotion Classes Creation}
\label{dataset}
For the current study, we collected, united (i.e. speech samples representatives of the same emotion were combined) and used speeches from 4 existing datasets: Emodb\footnote{\url{http://emodb.bilderbar.info/docu/}}, Emovo\footnote{\url{http://voice.fub.it/activities/corpora/emovo/index.html}}, Savee\footnote{\url{http://personal.ee.surrey.ac.uk/Personal/P.Jackson/SAVEE/}}, RAVDESS \cite{ravdess}. The speech files across all datasets are sampled at 48 kHz.

Emodb is a publicly available German speech database containing speech recordings of seven emotions: sadness, anger, happiness, fear, disgust, boredom and neutrality. The recordings were made by five men and five women, all actors, creating 10 statements for each emotion that was tested and evaluated. The entire dataset contains 535 speeches.

Emovo is a publicly available Italian speech database with seven emotions: happiness, sadness, anger, fear, disgust, surprise and neutral. Six Italian actors (three women and three men) generated 14 samples for each emotion. The entire dataset contains 588 speeches.

Savee is a British English language database of public speaking, made of speech recordings with seven emotions: happiness, sadness, anger, fear, disgust, surprise and neutral. The recordings were made by four English male actors, generating 15 samples for each emotion. The entire dataset contains 480 speeches.

Ravdess has been created by 24 authors: 12 men and 12 women with eight emotions: neutral, calm, happy, sad, angry, scared, disgusted, surprised. The entire dataset contains 1440 speeches.

Despite the fact that speeches across the datasets are sampled at the same frequency, they are structured in different and specific ways. Hence, we had to classify them appropriately and combine the same emotions from all datasets. The average length of speech samples is 3-4 seconds.

After the collection and classification of all emotions from the four different datasets, we got 10 classes of emotions: anger, anxiety, fear, boredom, calm, disgust, happiness, neutral, sadness, surprise. Due to the small number of $fear$ and $boredom$ emotions samples, we decided to add them respectively to $anxiety$ and $calm$ emotion classes. This was also motivated by the fact that, those speech samples have similar pitches and amplitudes. This results in 8 classes of emotion.

We chose to disregard gender information and to focus only on emotions. The final distribution of emotion classes with the numbers of speech samples of each class, are stated in Table \ref{class_emotion}.
\begin{table}
\begin{center}
\begin{tabular}{|c|c|}
\hline
Emotion class & Amount of data (\%) \\
\hline
Anger & 463 (15.22\%)\\
Anxiety & 405 (13.31\%)\\
Calm & 273 (8.97\%)\\
Disgust & 382 (12.55\%)\\
Happiness & 407 (13.37\%)\\
Neutral & 379 (12.45\%)\\
Sadness & 398 (13.09\%)\\
Surprise & 336 (11.04\%) \\
\hline
Total & 3043 (100\%)\\
\hline
\end{tabular}
\end{center}
\caption{\label{class_emotion} Distribution of emotion classes with the numbers of speech samples of each class.}
\end{table}

\section{Data Processing and Visualisation}
\label{data_processing}
\begin{figure}[t]
\begin{center}
\includegraphics[width=0.8\linewidth]{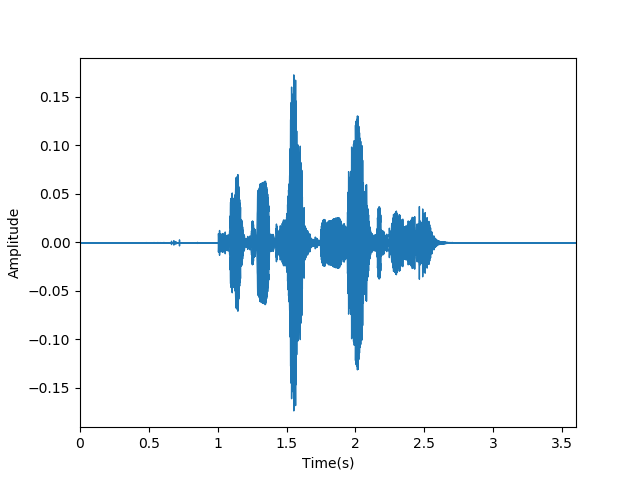}
\end{center}
   \caption{Speech sample visualisation using the $Librosa$ library}
\label{fig1}
\end{figure}
Figure \ref{fig1} shows the amplitudes of a random speech sample along the time measured in seconds (s). Speech signals are made up of amplitudes and frequencies. To get more information from our speech samples, we decided to map them into the frequency domain using the Fast Fourier Transformation (FFT) \cite{fourier_application,fourier}.
\begin{figure}[t]
\begin{center}
\includegraphics[width=0.8\linewidth]{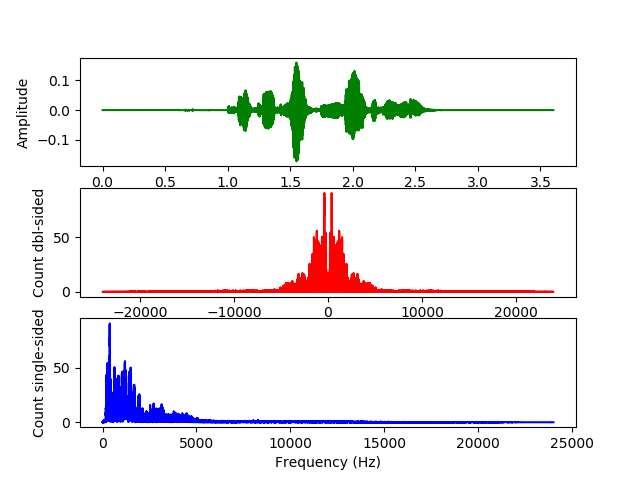}
\end{center}
   \caption{Speech sample characteristics using $Librosa$, and \textit{scipy.fft} libraries.}
\label{fig2}
\end{figure}

\begin{figure}[t]
\begin{center}
\includegraphics[width=0.8\linewidth]{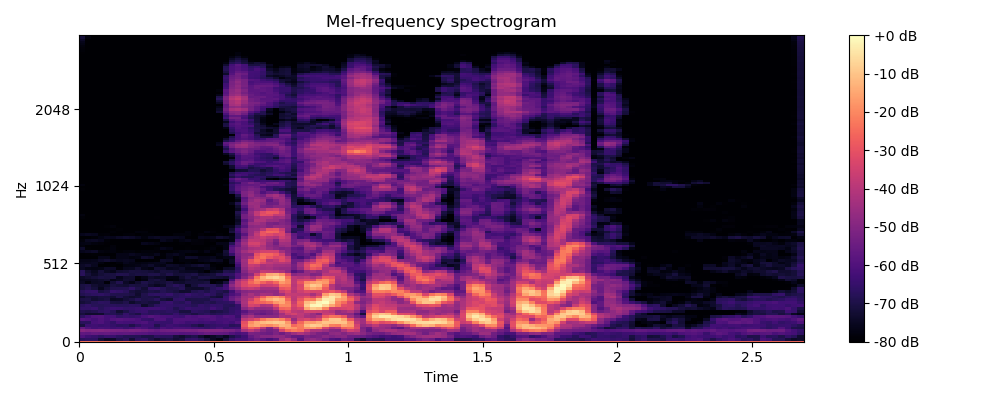}
\end{center}
   \caption{Speech sample converted into mel-spectrogram, with time, frequency and amplitudes information.}
\label{fig2_1}
\end{figure}

\begin{figure*}[htbp]
\begin{minipage}[b]{1\linewidth}
\centering
\subfloat[]{\includegraphics[width=7cm]{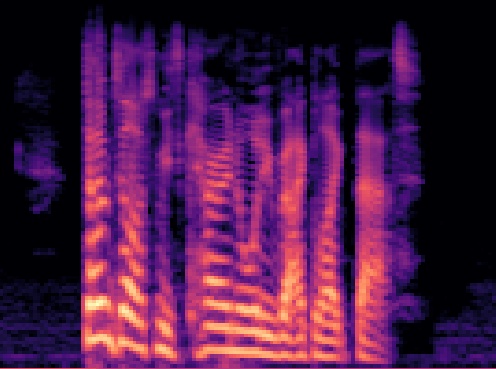}}\quad
\subfloat[]{\includegraphics[width=7cm]{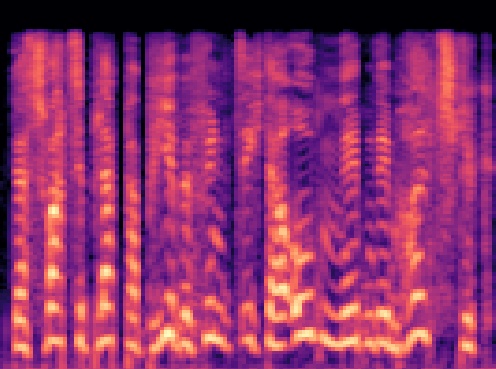}}\quad
\caption{\label{fig3} Mel-Spectrograms of speakers speaking a sentence with anger (a) and anxiety (b).}
\end{minipage}
\end{figure*}

Applying the FFT to a sample using the \textit{scipy.fft} and $librosa$ packages, gives Figure \ref{fig2}, which shows the initial speech plot, the double-side (negative and positive) FFT spectrum, and the positive FFT spectrum.

However, this is still lacking of time information. To remedy this, and make sure we preserve frequencies, time and amplitudes information about the speech samples, in reasonable and adequate range, we decided then to use mel-spectrograms.

In a mel-spectrogram\footnote{\url{http://www.glottopedia.org/index.php/Spectrogram}} the abscissa represents time, the ordinate axis represents frequency, and amplitudes are showed by the darkness of a precise frequency at a particular time: low amplitudes are represented with a light-blue color, and very high amplitudes are represented by dark red (see Figure \ref{fig2_1}).

There are two types of spectrograms: broad-band spectrograms and narrow-band spectrograms. In our study, we used narrow-band spectrograms because they have higher frequency resolution, and larger time interval for every spectrum than broad-band spectrograms: this allows the detection of \textit{very small} differences in frequencies. Moreover, they show individual harmonic structures, which are vibration frequency folds of the speech, as horizontal striations.

In Figure \ref{fig3} are provided examples of mel-spectrograms of speakers speaking a sentence with anger \ref{fig3}(a) and anxiety \ref{fig3}(b).
We converted all original speech signals to mel-spectrograms using the \textit{librosa melspectrogram}\footnote{\url{https://librosa.org/doc/main/generated/librosa.feature.melspectrogram.html}} module of the $librosa$ library, and transformed the SER task into a pattern recognition and image classification problem.

We used 512 as length of the FFT window, 512 as the hop-length (number of samples between successive frames) and a hanning windows size is set to the length of FFT window.

\paragraph{Dataset Splitting:}2434 images (80\%) have been allocated for the training phase. The remaining 609 (20\%) have been used as testing set, to evaluate the model performance. As a standard in the machine learning field, we then split the initial training into two subsets: training set (80\% - 1947 images) and validation set (20\% - 487 images). To increase the size of the final training set, we applied a very common technique that has shown much success in recent years with CNNs: data augmentation \cite{data_aug1, data_aug2}.
\paragraph{Data Augmentation:}
Using the algorithm \ref{algo}, we implemented the data augmentation (DA) method, generating 20 additional pictures from every original input picture of our training set. To each input image (mel-spectrogram) object, we applied width and height shifting, zooming and horizontal flipping, using the \textit{keras.preprocessing.image} module from the $Keras$ deep learning framework\footnote{\url{https://keras.io/api/preprocessing/image/}}. After the DA, our training set size increased to \textit{40887} images (mel-spectrograms).
\raggedbottom

\begin{figure*}
\begin{center}
\includegraphics[width=\linewidth]{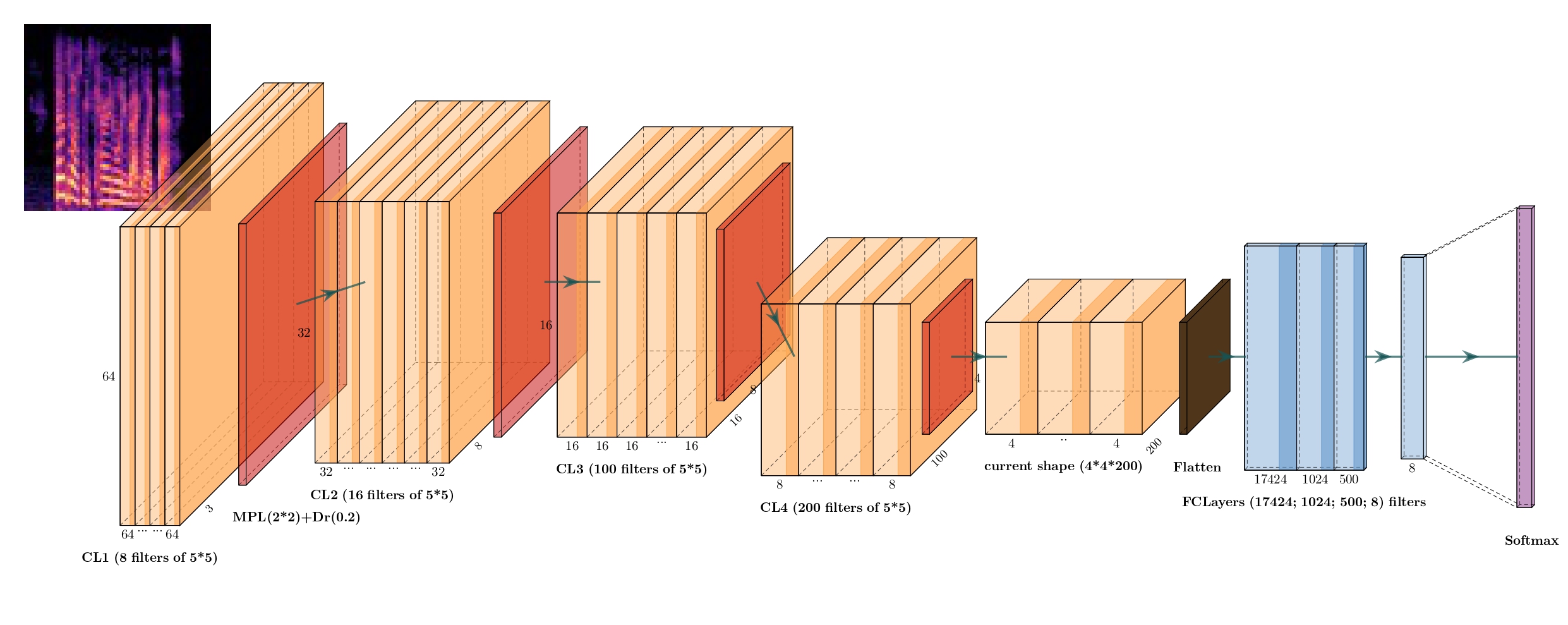}
\end{center}
   \caption{FSER Model Architecture with: 1) CL:  Convolutional Layer + Relu Activation function, 2) MPL: Maximum 2D Pooling Layer, 3) Dr: Dropout, and 4) FCL: Fully-Connected (Dense) Layer.}
\label{fig4}
\end{figure*}

\begin{figure}[t]
\begin{center}
\includegraphics[width=0.8\linewidth]{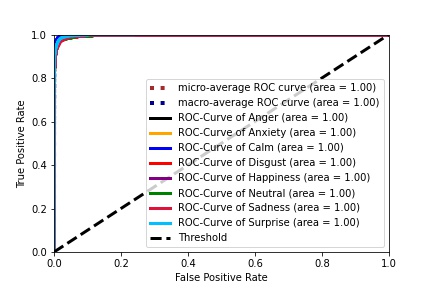}
\end{center}
   \caption{FSER ROC-AUC on 8-emotions Recognition and Classification}
\label{fig5}
\end{figure}

\begin{table}[]
\resizebox{\columnwidth}{!}{
    \centering
    \begin{tabular}{llllll}
    \toprule
        \textbf{} & Precision & Recall & F1-Score &Support&AUC-Score\\
    \midrule
        Anger&96&96&\textbf{96}&91&\textbf{99.84}\\
        Anxiety&94&\textbf{97}&95&80&\textbf{99.85}\\
        Calm&93&\textbf{97}&95&54&\textbf{99.92}\\
        Disgust&\textbf{97}&92&94&77&\textbf{99.72}\\
        Happiness&\textbf{95}&93&94&83&\textbf{99.75}\\
        Neutral&94&\textbf{94}&94&77&\textbf{99.74}\\
        Sadness&\textbf{95}&94&94&78&\textbf{99.76}\\
        Surprise&94&\textbf{95}&94&69&\textbf{99.82}\\
        \hline
        accuracy & - &  & \textbf{95} & 609&-\\
        macro avg & 95 & 95 & \textbf{95} & 609&\textbf{99.79}\\
        weighted avg & 95 & 95 & \textbf{95} & 609&\textbf{99.79}\\
    \hline
    \bottomrule
    \end{tabular}
    }
    \caption{Classification Report of the Predictions of FSER on the Testing set}
    \label{class_rep}
\end{table}

\begin{table*}[!t]
\begin{center}
\resizebox{\textwidth}{!}{
    \centering
\begin{tabular}{|c||c|c|c|c|c|c|c|c|}
 \hline
 &\multicolumn{8}{ c|}{Predicted Emotions} \\
 \hline
 Expected Emotions $\downarrow$&Anger&Anxiety&Calm&Disgust&Happiness&Neutral&Sadness&Surprise\\
 \hline
 Anger&\textbf{96.92}&0.31&0.16&0.42&1.15&0.42&0.37&0.26\\
 Anxiety&0.53&\textbf{95.46}&0.47&0.29&1.12&0.53&0.94&0.65\\
 Calm&0.00&0.18&\textbf{96.59}&0.00&0.18&1.97&0.81&0.27\\
 Disgust&0.88&1.19&1.00&\textbf{91.49}&0.56&2.06&1.44&1.38\\
 Happiness&1.72&0.98&0.40&0.40&\textbf{93.16}&1.03&0.86&1.44\\
 Neutral&0.37&0.50&1.43&0.50&0.50&\textbf{95.45}&1.06&0.19\\
 Sadness&0.41&0.94&1.00&0.35&0.23&0.59&\textbf{95.84}&0.64\\
 Surprise&0.43&1.58&0.14&0.50&0.65&0.14&1.08&\textbf{95.46}\\
 \hline
\end{tabular}
    }
\caption{\label{confusion_matrix}Normalized Confusion Matrix of FSER on the SER Task}
\end{center}
\end{table*}

\section{Model Architecture, Training and Results}
\label{results}
\subsection{Model Architecture and Training}
Figure \ref{fig4} presents the architecture of FSER. The input is an RGB image of shape (64, 64, 3). Our model is made of four blocks of convolutional layers (CLs) with reLU activation \cite{relu}, maximum 2D pooling layers (MPLs) \cite{max_pooling} and Dropouts \cite{francois_cholet} interposed between them.
\begin{table*}[!t]
\begin{center}
\resizebox{\textwidth}{!}{
    \centering
\begin{tabular}{|c||c|c|c|c|c|c|c|c|}
 \hline
 &\multicolumn{4}{ c|}{Speech Dataset}&\multicolumn{2}{c|}{Features}&\multicolumn{2}{c|}{Accuracies (\%)} \\
 \hline
 Relevant related works $\downarrow$&Emodb&Emovo&Savee&Ravdess&Mel-Spectrogram (*)&MFCC (**)&*&**\\
 \hline
 Qayyum \etal \cite{cnn_spec1}&-&-&+&-&-&+&-&83.61\\
 Kannan, V. and Haresh, R. \cite{speech_1}&-&-&-&+&+&+&86&53\\
  Zhao \etal \cite{cnn_spec2}&+&-&-&-&+&-&95.33&-\\
\textbf{Our work (Emodb)}&\cmark&-&-&-&\cmark&-&\textbf{97.74}&\xmark\\
\textbf{Our work (Savee)}&-&-&\cmark&-&\cmark&-&\textbf{99.10}&\xmark\\
\textbf{Our work (Ravdess)}&-&-&-&\cmark&\cmark&-&\textbf{98.67}&\xmark\\
 \hline
\end{tabular}
    }
\caption{\label{table_comparison}Comparison of our work to other relevant related works on Emodb, Savee and Ravdess datasets.}
\end{center}
\end{table*}
The output of the four CLs is flattened and fed into a series of three fully-connected layers, which is finally fed into a softmax activation unit \cite{softmax} to output the probability of the input to belong to each class of emotion.

\paragraph{Hyper-parameters and Training:}We used a batch size of 64, a learning rate of 0.001 and Stochastic Gradient Descent \cite{sgd,sgd1} as optimizer. The loss function is the categorical cross-entropy.

Our hyper-parameters (stated above), including the number of layers, the number of filters in each layer, the dropout probability, the size of the kernel, padding and stride, have been selected after many manual fine-tuning trials.

FSER has been trained for 400 epochs, during 8 days on a 16GB Tesla K80 GPU, using the platform \textit{PaperSpace}.

\subsection{Results and potential use case}
Tables \ref{class_rep} and \ref{confusion_matrix} present respectively the classification report of FSER on the SER task and the normalized confusion matrix. As shown in Tables \ref{class_rep} and \ref{confusion_matrix}, and in Figure \ref{fig5}, FSER has a nearly-perfect score for every emotion class.

Additionally, Table \ref{table_comparison} shows the comparison of our work to other works done on the SER task, with the datasets of interest. We can interpret Table \ref{table_comparison} as follow:
\begin{itemize}
    \item On Emodb, our FSER (97.44) outperformed the mel-spectrograms-based approach proposed by Zhao \etal\cite{cnn_spec2} (95.53).
    \item On Savee, our FSER (99.10) outperformed the traditional MFCCs-based approach proposed by Qayyum \etal \cite{cnn_spec1} (83.61)
    \item On Ravdess, our FSER (98.67) outperformed both mel-spectrograms and traditional MFCCs approaches proposed by Kannan, V. and Haresh, R. \cite{speech_1} (86 and 53 respectively).
\end{itemize}
To test the robustness of FSER, we collected real-life audio samples from French, Fon, and Igbo speakers. Each of them was asked to imitate as much as they can those emotions, without providing any guidelines. In total, we had 24 audios (3 samples per emotion class), out of which FSER predicted correctly 20 audios. The 4 misclassed audios were found to belong to the classes $fear$ and $boredom$ that we had to respectively add and consider as $anxiety$ and $calm$ classes. Recalling that FSER has been trained in English, German, and Italian, its performance on these real-time audios, shows that FSER stays reliable, independently of the language, sex identity, and any other external factor.

\paragraph{Potential use case:} For humans, emotions constitute an important factor to personal and global health. Being emotionally unstable can affect not only mental wellness but also physical health. During this pandemic, many health reports demonstrated a significant increase in feelings of loneliness and emotional instability leading to and increased rate of suicide for example. Being emotionally healthy is hence crucial for our well-being. As AI-based systems are nowadays helping doctors to quickly diagnose diseases such as cancers, FSER-like systems could also be useful to doctors, particularly psychologists, in identifying the emotional state of their patient. It is no news that some people have trouble expressing their emotions with words, so FSER brings a potential solution to help physicians to better understand their patients, and provide them more efficient treatments. They could also be useful to organizations providing mental care and support to families, clients, or communities at large.
\section{Conclusion}
In the current study, we evaluated the abilities of CNNs to effectively recognize and classify emotions from speech data. We introduced FSER, which outperformed all models that have been introduced for SER task. We showed that FSER is reliable, with no regards to the language spoken, the sex identity of the individual, or any other external factor. FSER-like systems could be beneficial to the healthcare system, as they can help to better understand patients, to provide them better and quicker treatments. It is worth mentioning the limitations of this work, notably the data augmentation we had to perform to deal with the limited number of samples.

{\small
\bibliographystyle{ieee_fullname}
\bibliography{egbib}

\begin{thebibliography}{10}\itemsep=-1pt

\bibitem{cnn_spec1}
A.~B. {Abdul Qayyum}, A. {Arefeen}, and C. {Shahnaz}.
\newblock Convolutional neural network (cnn) based speech-emotion recognition.
\newblock In {\em 2019 IEEE International Conference on Signal Processing,
  Information, Communication Systems (SPICSCON)}, pages 122--125, 2019.

\bibitem{psy_3}
Daniel~C. Albas, Ken~W. McCluskey, and Cheryl~A. Albas.
\newblock Perception of the emotional content of speech: A comparison of two
  canadian groups.
\newblock {\em Journal of Cross-Cultural Psychology}, 7(4):481--490, 1976.

\bibitem{fourier_application}
B. Boashash.
\newblock {\em Time-Frequency Signal Analysis and Processing: A Comprehensive
  Reference}.
\newblock Oxford: Elsevier Science, 2003.

\bibitem{fourier}
R.~N Bracewell.
\newblock {\em The Fourier Transform and Its Applications}.
\newblock Boston: McGraw-Hill, 2000.

\bibitem{francois_cholet}
François Chollet.
\newblock {\em Deep Learning with Python}.
\newblock Manning, Nov. 2017.

\bibitem{psy_2}
Pablo Fossa, Raymond~Madrigal Pérez, and Camila~Muñoz Marcotti.
\newblock The relationship between the inner speech and emotions: Revisiting
  the study of passions in psychology.
\newblock {\em Human Arenas}, 2020.

\bibitem{softmax}
Ian Goodfellow, Yoshua Bengio, and Aaron Courville.
\newblock {\em Deep Learning}.
\newblock MIT Press, 2016.
\newblock \url{http://www.deeplearningbook.org}.

\bibitem{psy_4}
Tom Johnstone.
\newblock The effect of emotion on voice production and speech acoustics, Aug
  2017.

\bibitem{ravdess}
Steven~R. Livingstone and Frank~A. Russo.
\newblock {The Ryerson Audio-Visual Database of Emotional Speech and Song
  (RAVDESS)}, Apr. 2018.
\newblock {Funding Information Natural Sciences and Engineering Research
  Council of Canada: 2012-341583 Hear the world research chair in music and
  emotional speech from Phonak}.

\bibitem{psy_6}
Sydney Lolli, Ari Lewenstein, Julian Basurto, Sean Winnik, and Psyche Loui.
\newblock Sound frequency affects speech emotion perception: results from
  congenital amusia.
\newblock {\em Frontiers in Psychology}, 6:1340, 2015.

\bibitem{relu}
Vinod Nair and Geoffrey~E. Hinton.
\newblock Rectified linear units improve restricted boltzmann machines.
\newblock In Johannes Fürnkranz and Thorsten Joachims, editors, {\em ICML},
  pages 807--814. Omnipress, 2010.

\bibitem{psy_1}
Kislova O.O. and Rusalova M.N.
\newblock Perception of emotions in speech. a review of psychological and
  physiological research.
\newblock {\em Usp Fiziol Nauk}, 2013.

\bibitem{data_aug1}
Luis Perez and Jason Wang.
\newblock The effectiveness of data augmentation in image classification using
  deep learning, 2017.

\bibitem{sgd}
Herbert Robbins and Sutton Monro.
\newblock A stochastic approximation method.
\newblock {\em Ann. Math. Statist.}, 22(3):400--407, 09 1951.

\bibitem{sgd1}
Sebastian Ruder.
\newblock An overview of gradient descent optimization algorithms, 2016.

\bibitem{psy_5}
K.R. Scherer.
\newblock Personality, emotion, psychopathology and speech.
\newblock In HOWARD GILES, W~PETER ROBINSON, and PHILIP~M SMITH, editors, {\em
  Language}, pages 233--235. Pergamon, Amsterdam, 1980.

\bibitem{speech_1}
Kannan Venkataramanan and Haresh~Rengaraj Rajamohan.
\newblock Emotion recognition from speech, 2019.

\bibitem{data_aug2}
Sebastien~C. Wong, Adam Gatt, Victor Stamatescu, and Mark~D. McDonnell.
\newblock Understanding data augmentation for classification: when to warp?,
  2016.

\bibitem{max_pooling}
Haibing Wu and Xiaodong Gu.
\newblock Max-pooling dropout for regularization of convolutional neural
  networks, 2015.

\bibitem{cnn_spec2}
Jianfeng Zhao, Xia Mao, and Lijiang Chen.
\newblock Speech emotion recognition using deep 1d and 2d cnn lstm networks.
\newblock {\em Biomedical Signal Processing and Control}, 47:312--323, 2019.

\end{thebibliography}
}
\newpage
\label{algorithms}
\begin{lstlisting}[language=Python, caption=Algorithm for the generation of new samples for Data Augmentation, label=algo]
from keras.preprocessing.image import *
import numpy as np

for image in images:
    # images is the list of all images
    image = load_img(image)
    image = img_to_array(image)
    image = np.expand_dims(image, 0)

    # construct the image data 
    # generator for data augmentation
    aug = ImageDataGenerator(*args)
    
    # generating images
    imageGen = aug.flow(image, *args)
    total = 0
    # loop over image data generator
    for image_generated in imageGen:
        total += 1
        if total == 20:
            break
\end{lstlisting}

\end{document}